\documentstyle[preprint,aps,epsfig]{revtex}
\begin{document}

\tighten
\draft
\preprint{
\vbox{
\hbox{\today}
\hbox{Tashkent}
}}
\newcommand{\ci}[1]{\cite{#1}}
\newcommand{\lab}[1]{\label{#1}}
\newcommand{\re}[1]{(\ref{#1})}
\newcommand{\bfr}{\begin{flushright}}
\newcommand{\bfl}{\begin{flushleft}}
\newcommand{\efl}{\end{flushleft}}
\newcommand{\efr}{\end{flushright}}
\newcommand{\bc}{\begin{center}}
\newcommand{\ec}{\end{center}}
\newcommand{\be}{\begin{equation}}
\newcommand{\ee}{\end{equation}}
\newcommand{\bea}{\begin{eqnarray}}
\newcommand{\eea}{\end{eqnarray}}
\newcommand{\ba}{\begin{array}}
\newcommand{\ea}{\end{array}}
\newcommand{\edc}{\end{document}}
\newcommand{\ul}{\underline}
\newcommand{\ri}{\rightarrow\infty}
\newcommand{\li}{\leftarrow\infty}
\newcommand{\ra}{\rightarrow}
\newcommand{\la}{\leftarrow}
\newcommand{\ds}{\displaystyle}
\newcommand{\dsf}{\displaystyle\frac}
\newcommand{\dt}{\Delta{t}}
\newcommand{\il}{\int\limits}
\newcommand{\pal}{\partial}
\newcommand{\xxx}{{\it{X}}}
\newcommand{\bone}{{\bf 1}}

\title{Complex diffusion Monte-Carlo method:\\
test by the simulation of
the $2D$ fermions}

\author{B. Abdullaev$^1$, M. Musakhanov$^1$ and A. Nakamura$^{2}$ }
\address{$^1$
 Theoretical  Physics Dept, Uzbekistan National University,\\
 Tashkent 700174, Uzbekistan\\
e-mail: bah$\_$abd@iaph.silk.org, yousuf@iaph.silk.org}
\address{$^2$
RIISE, Hiroshima University, Higashi-Hiroshima 739-8521, Japan \\
e-mail: nakamura@riise.hiroshima-u.ac.jp}

\maketitle

\begin{abstract}
On the base of the diffusion Monte-Carlo method we  develop the method
allowing to simulate the quantum systems with complex wave function.
The method is exact and there are no approximations on
the simulations of the module and the phase of the system's wave
function. In our method
 averaged  value of any  quantity  have no direct contribution
from the phase of distribution function but only
from the phase of the Green function of diffusion equation.

We test the method by the simulations of
the ground state of fermions in two-dimensional parabolic well.
Anyons are used for the representation of the
two-dimensional (2D) fermions. We
vary the number of fermions from two to ten and find a good
agreement of the numerical results with analytical ones for the
numbers of the particles $N > 4$.

\end{abstract}

\section{Introduction}

There are several objects of the  quantum mechanics where the investigated
system's wave function is essentially complex(having imaginary part). The examples of these
objects are a two-dimensional system of electrons in the external
uniform magnetic field when vector potential has a central-symmetric
form and system of anyons. The complexity of wave function does not give
a possibility for the simulations of such kind of systems by using
well-known  Green Function Monte-Carlo method (see review \ci{9}).
This method essentially demand
the reality of the system's wave function which considered
as probability weight in during stochastic process.

There had been undertaken  the several attempts \ci{10,11,12} for the
construction a method for the simulation of quantum systems with complex
wave function.
Authors of the works \ci{10,13} suggested
to take as a probability weight the module of
the complex distribution function.
All quantities are calculated
in \ci{10,13} by averaging over this complex distribution function.
The phase of this function $\gamma $ is taken accordingly
$ \gamma = \sum \gamma_i$, where $\gamma_i $
is the phase at the $i$-time step of the stochastic process.

Another quantum Monte Carlo method using algorithm without
branching for the simulating of
complex problems was developed in \ci{12}.
The main trouble of this approach
was an increasing of the statistical error as a function of
the whole time of the simulation.

In the previous paper \ci{AMN1} it was suggested new
Complex Diffusion Monte-Carlo (CDMC) method
allowing to simulate the quantum systems
with complex wave function.
Our method is rather close to the fixed - phase Diffusion Monte-Carlo method
developed for the  simulation of the two-dimensional electrons in
the magnetic field \ci{11}. In this work there had been an exact simulation
only of the module of system's wave function.
The phase of the
wave function was considered as fixed and equal to phase of the Laughlin's
wave function \ci{14}.    In contrast,  our method
include also the simulation of the phase factor of the wave function.

  The basic difficulty of the numerical simulation of the fermions
is essentially the same as
for the systems with the complex wave function.
Their wave function can change the sign and therefore
can not be used as the probability weight in the simulation process.

For the simulation of the continuum (not on the lattice) fermionic systems
it were developed widely  used  fixed node Monte-Carlo method
\ci{1}( see also reviews \ci{9,15}) and recently proposed
constrained path Monte-Carlo method \ci{16}. In
these both methods it was assumed the restriction on the random walks
connected with the uncertainties in the space localization of the wave
function node surfaces. The comparison of these methods was done in \ci{17}.

In the $2D$ space we have unique tool --  so-called
anyons, particles (bosons or fermions) with additional
gauge interaction which provide needed  statistical properties \ci{3,4}.
So, by tuning the coupling constant we may arrive to the fermions
starting from the bosons.
It looks very attractive to apply the anyons for the simulation of
the $2D$ fermions, because statistical vector potential
 that gives anyonic property is relatively smooth and continuous \ci{4,5}.
The remaining main problem is that the wave function of anyonic system
is essentially complex, i.e. it contains an imaginary part.

The CDMC method was successfully tested by the simulation
of ground state of one electron in magnetic field \ci{AMN1}.
In the present
work we briefly outline the idea of CDMC (the details of the algorithm
one can find in \ci{AMN1}) and provide a further test it
by simulating the
ground state  of fermions in $2D$ parabolic well and comparing with
well known analytical answer for the ground state energy as a function of
the number of fermions.
The simulation is done
for the fermionic systems with the number of particles from two to ten.
It was found a good
agreement of the numerical results with analytical ones for the
numbers of the particles great than four.
This observation
of the fermionic simulation allows
us to hope that the CDMC is a good tool  for the simulation
of $2D$ fermion systems.

\section{Model and  CDMC method for the simulation}

Hamiltonian of anyons in the $2D$ parabolic well has a form:
\be
\hat H=\dsf{1}{2M}\ds\sum_{i=1}^N(\vec p_i+\vec A(\vec r_i))^2+
\dsf{Mw_o^2}{2}\ds\sum_{i=1}^N\vec r_i^2.
\lab{1}
\ee
Here $M$ is the mass of particle,  $\vec p=-i\hbar \vec
\nabla$, where
$\vec \nabla=\vec i \dsf{\pal}{\pal x}+ \vec j \dsf{\pal}{\pal y}$,
$w_o$ stands for characteristic frequency of free
particles in $2D$ parabolic well and $\vec r_i$ is radius vector of
$i$-th particle. The number of particles is $N$.

Vector potential for anyons $\vec A(\vec r_i)$ \ci{4,5} in (1) is
\be
\vec A(\vec r_i)=\hbar\nu\ds\sum_{j>i}^N\dsf{\vec z \times\vec r_{ij}}
{|\vec r_{ij}|^2}.
\lab{2}\ee
Here $\vec z$ is unit vector perpendicular to $2D$ plane and $\nu$
characterizes the form of fractional statistics ( spin of the anyon ).
In this work we want to employ a bosonic representation of anyons, therefore
$\nu=0$ corresponds to noninteracting bosons and $\nu=1$ is
the case of our interest - free fermions.

   In CDMC a Monte Carlo simulation
is carried out with
the complex distribution function
\be
f(\vec R,t)=\Psi_T^*(\vec R)\Psi(\vec R,t),
\lab{3}\ee
where $\vec R$ stands for the coordinates of all particles,
$\Psi_T(\vec R)$ is a trial wave function
and the wave function
$\Psi(\vec R,t)$ satisfies a Schr\"{o}dinger equation
with imaginary time (expressed in $\hbar$ units)
\be
-\dsf{\pal \Psi(\vec R,t)}{\pal t}=(\hat H-E_T)\Psi(\vec R,t).
\lab{4}\ee
\noindent

With a suitable choice of a trial energy $E_T$
it is possible to arrange the convergence
    $\Psi(\vec R,t)\ra\Psi_0(\vec R)$
- an exact stationary wave function of ground state of Hamiltonian \re{1}
at $t\ri$ \ci{1}.

  For a bosonic representation of anyons we take
(conjugated) trial wave function in the form:
\be
\Psi_T^*(\vec R)=\prod_{i=1}^N\Psi_T^*(\vec r_i).
\lab{5}\ee
  Good choice of single particle trial wave function
$\Psi_T^*(\vec r_i)$ ( see below ) as  energy $E_T$ must  provide
a  convergence of simulation process to exact ground state.

   The distribution function $f\equiv f(\vec R,t)$ satisfies a diffusion
equation:
\bea
-\dsf{\pal f}{\pal t}&=& -D\sum_{i=1}^N\Delta_if+D\sum_{i=1}^N\vec\nabla_i
(f \mathrm{Re} \vec F_Q(\vec r_i))+
\nonumber \\
&& +i\ds\sum_{i=1}^N[\vec\nabla_i(Df \mathrm{Im} \vec F_Q(\vec r_i))-
\dsf{\hbar}{M}\vec A(\vec r_i)\vec \nabla_i f]+(E_L(\vec R)-E_T)f.
\lab{6}
\eea
Here $D=\dsf{\hbar^2}{2M}$, $\Delta_i=\vec\nabla_i^2$.
When a time step integration of diffusion equation \re{6},
$\tau$, goes to zero, then Green function for this equation  has a form:
\bea
G(\vec R,\vec R\ ';\tau)&=&\dsf{\exp[D\tau\ds\sum_{i=1}^N
\vec A_Q^2(\vec r_i,\vec r_i\ ')]}{(4\pi D\tau)^N}\times
\nonumber \\
&&\times\exp\left[-\dsf{\ds\sum_{i=1}^N(\vec r_i-\vec r_i\ '-D\tau
\mathrm{Re} \vec F_Q(\vec r_i\ '))^2}{
4D\tau}\right]\exp\left[-\tau(E_L(\vec R)-E_T)\right]\times
\nonumber \\
&&\times
\exp\left[i\ds\sum_{i=1}^N\vec A_Q(\vec r_i,\vec r_i\ ')
(\vec r_i-\vec r_i\ '-D\tau \mathrm{Re} \vec F_Q(\vec r_i\ '))\right].
\lab{7}
\eea
In the expressions \re{6} and \re{7}
\be
\vec F_Q(\vec r_i)=2\Psi_T^{*-1}(\vec r_i)\vec\nabla_i\Psi_T^{*}(\vec r_i),
\lab{8}\ee
\be
E_L(\vec R)=\Psi_T^{*-1}(\vec R)\hat H'\Psi_T^{*}(\vec R),
\lab{9}\ee
\be
\hat H'=\dsf{1}{2M}\ds\sum_{i=1}^N(\vec p_i-\vec A(\vec r_i))^2+
\dsf{Mw_o^2}{2}\ds\sum_{i=1}^N\vec r_i^2.
\lab{10}\ee
In general case $\Psi_T^*(\vec r_i)$ is a complex wave function and
\be
\vec F_Q(\vec r_i)=\mathrm{Re} \vec F_Q(\vec r_i)+i \mathrm{Im}
\vec F_Q(\vec r_i).
\lab{11}\ee
In \re{7} (see also \ci{AMN1}) we have introduced a new quantity
\be
\vec A_Q(\vec r_i,\vec r_i\ ')=\dsf{1}{2}\mathrm{Im} \vec F_Q(\vec r_i\ ')-
\dsf{1}{\hbar}\vec A (\vec r_i).
\lab{12}\ee
In the expressions \re{7} and \re{12}
vectors $\vec R$ and  $\vec r_i$ correspond to time point
$t+\tau$ and vectors $\vec R\ '$,  $\vec r\ '$ to time point $t$.

One can see that the Green function $G(\vec R,\vec R\ ';\tau)$ \re{7}
as and distribution function $f$ in \re{3} is a complex.
These both quantities are satisfying to the usual integral relation
\be
f(\vec R,t+\tau)=\int d\vec R\ ' G(\vec R,\vec R\ ';\tau)f(\vec R\ ',t).
\lab{a15}\ee
From relation \re{a15} it is following that the module and phase of the
distribution function at consequent time point are determining by
the module and the phase of the Green function and of ones of the
distribution function at previous time point of integration of
diffusion equation \re{6}.

As the energy $E_L(\vec R)$ is a complex,  real and imaginary
part of  $E_L(\vec R)$ contribute separately to ones of the Green function.
Therefore the expression of the last two exponents in \re{7} has a form:
\bea\ba{c}
\exp\left[-\tau(\mathrm{Re}E_L(\vec R)-E_T)\right]\times\\
\times
\exp\left[i\ds\sum_{i=1}^N\vec A_Q(\vec r_i,\vec r_i\ ')
(\vec r_i-\vec r_i\ '-D\tau \mathrm{Re} \vec F_Q(\vec r_i\ '))-
i\tau\mathrm{Im}E_L(\vec R)\right].
\lab{a12}
\ea\eea

The mean value of the some quantity
$B(\vec R(t))$ on the complex distribution function $f(\vec R,t)$ at the
time $t$ of the simulating process is calculated by the formula (see
\ci{10,13})
\be
<B(t)>=\dsf{\ds\sum_{i=1}^M\exp[i\gamma(\vec R_i(t)]B(\vec R_i(t))}{
\ds\sum_{i=1}^M\exp[i\gamma(\vec R_i(t)]}.
\lab{a1}\ee
Here $\vec R_i(t)$ is the coordinates of all particles from the $i$-th
configuration of the
 $M$ - configurations at time $t$.
The coordinates $\vec R_i(t)$ of particles
in \re{a1} are weighted with probability $|f(\vec R_i,t)|$
and the quantity $\gamma(\vec R_i(t))$ is a phase of the distribution function
$f(\vec R_i,t)$.

From integral expression \re{a15} it is seen that phase of the distribution
function $\gamma(\vec R_i(t+\tau))$ connected with the phase of the
Green function $\gamma_G(\vec R_i(t+\tau),\vec R_j(t))$ and the phase
$\gamma(\vec R_j(t))$
of the distribution function by the relation:
\be
\gamma(\vec R_i(t+\tau))=\gamma_G(\vec R_i(t+\tau),\vec R_j(t))+
\gamma(\vec R_j(t))\nonumber.
\ee
Here  $\vec R_j(t)$ is $j$- th configuration at time $t$.
So, we can write
\bea\ba{c}
<B(t+\tau)>=\dsf{\ds\sum_{i=1}^M e^{i\gamma_G(\vec R_i(t+\tau), \vec R_j(t))+
i\gamma(\vec R_j(t))}B(\vec R_i(t+\tau))}{
\ds\sum_{i=1}^Me^{i\gamma_G(\vec R_i(t+\tau), \vec R_j(t))+
i\gamma(\vec R_j(t))}}=\\
=\dsf{\ds\sum_{i=1}^Me^{i\gamma_G(\vec R_i(t+\tau), \vec R_j(t))}B(\vec R_i(t+\tau))}{
\ds\sum_{i=1}^Me^{i\gamma_G(\vec R_i(t+\tau), \vec R_j(t))}}\lab{a23}.
\ea\eea

The expression \re{a23} is hold, because all quantities
in numerator and denominator are weighted with probability
$|f(\vec R_i(t+\tau),t+\tau)|$, i.e. at time moment
$t+\tau$ with new configurations  $\vec R_i(t+\tau)$.
Hence, a mean quantity $<B(t+\tau)>$ is determined only by phase
 $\gamma_G$ of the Green function.  It is a key conclusion, providing
 the calculation of mean quantities in our CDMC.

Next, by taking into account the expression for the simulation of the random
displacement for the particle
$\vec r_i=\vec r_i\ '+D\tau\mathrm{Re} \vec F_Q(\vec r_i\ ')+
2(D\tau)^{1/2}\chi$ (see \ci{1} and \ci{AMN1}), where $\chi$ is a gaussian
random number, the expressions for $\vec F_Q(\vec r_i)$
\re{8}, for $E_L(\vec R)$ \re{9} and for $\vec A_Q(\vec r_i,\vec r_i\ ')$
\re{12} and  also the expression \re{a12}, one can show that at
$\tau\rightarrow 0$ the phase of the Green function $\gamma_G$ tends to zero
as $\tau^{1/2}$. This property of $\gamma_G$ together with the formula
\re{a23}  allows CDMC successfully  simulate all mean quantities.

  The simulation process starts with mean field approximation of Fetter,
Hanna and Laughlin \ci{7}. In this approximation the  mean
field $\overline{\vec {A}}$ generated by the average density $\rho$ of the particles:
\be
\overline {\vec{A}}(\vec r)=\rho\pi\hbar\nu(\vec z\times\vec r) =
\dsf{1}{2}\vec B\times\vec r.
\lab{13}\ee
Here $\vec B=2\pi\rho\hbar\nu\vec z$ is fictitious uniform mean magnetic
field that defines corresponding magnetic length
$a_H=\sqrt{\hbar/B}$
and cyclotron frequency $\omega_c=B/M$. Particles move in this
mean magnetic field. Energy
spectrum of Hamiltonian \re{1} with $\overline {\vec A}$ and without $2D$
parabolic well is Landau levels. We have to simulate anyonic ground
state. Therefore we take a wave function $\Psi_T^*(\vec r_i)$ in the
following form:
\be \Psi_T^*(\vec
r_i)=C\exp\left(-\alpha\dsf{(x_i^2+y_i^2)}{2R_o^2}\right)
\exp\left(-\dsf{(x_i^2+ y_i^2)}{4a_H^2}\right).
\lab{14}\ee
Here C is a normalization constant, $\alpha$ is numerical parameter and
$R_o=(\hbar/Mw_o)^{1/2}$. In this expression first exponential function
with $\alpha=1$ corresponds to the ground state wave function of
particle in $2D$ parabolic well
and second one to the ground state of Landau levels.

Average density of particles $\rho$ is given by
$\rho=1/\pi r_o^2$  where $r_o$ is the mean distance between particles
in system. We assume that $r_o$ equals to $R_o$.
The reason of this is that the interaction between two anyons at any non-zero
$\nu$ has repulsive character and the area $S_n$ that occupied by free
particle at $n$-th $2D$ oscillator's level is $S_n=n\pi R_o^2$. Last
relation for
$S_n$ follows by using virial theorem for the quantum energetic states
for the free particle in $2D$ parabolic well.

  By substituting expression $B$ into expression $a_H$ and taking account
that $\rho=1/{\pi r_o^2}$, we find the following relation between
$a_H$ and $r_o$: $a_H=\sqrt{1/2\nu}\,r_o$. By similar way one can
show that $\hbar\omega_c/2=\nu\hbar\omega_o$.

  It is natural to use $\hbar \omega_c/2$ as energy unit,
$a_H$ as length unit and $2/\hbar\omega_c$ as time unit $\tau$.
In these units the last term of the Hamiltonian \re{1} has a coefficient
$1/4{\nu}^2$ and the argument of the first exponential function in
\re{14} has a coefficient $\alpha/4\nu$.  By using above relations
between $\hbar\omega_c/2$ and $\hbar\omega_0$ and between $a_H$ and
$r_0$ it is easily to express the energies in terms of $2D$
oscillator quanta $\hbar\omega_0$ and the lengths in terms of $r_0$.

\section{Discussion of the results}

  The simulation of the ground state of the systems of several fermions
in $2D$ parabolic well was performed.
By setting the coupling $\nu=1$ in
\re{2} we get fermions from anyons.
This value of $\nu$ is fixed below.
 As in \ci{1}, the total simulation time $t$ is
divided into several (in our case we took 40) time blocks
$\Delta t$.
In every time block $\Delta t$
initial number of configurations $N_c$
is chosen equal to 1000 and number of time steps $\tau$
equal to 30.

 We simulate the energy $E_{SIM}$, mean radius $\overline r$ and mean
square of radius of $\overline {r^2}$ of the
fermionic systems. The calculation of these quantities are performed
by the formula \re{a23}. Here we take into account that the energy $E_L(\vec R)$
has as a real part $\mathrm {Re} E_L(\vec R)$, and also an imaginary part
$\mathrm {Im} E_L(\vec R)$.
The number of fermions is varied from two to ten.

The $\tau$  dependence of the energy $E_{SIM}$
has a general structure $A_0 + A_1 \tau^{1/2}$ \ci{1} where $A_0$ and $A_1$
are the numerical constants.

We have obtained  that at a very small $\tau$
the mean deviation of  $E_{SIM}$ larger than the fluctuations of $E_{SIM}$.
At large $\tau$ further increasing of it leads to the increasing of
the fluctuation of $E_{SIM}$ and thus to the growing of the population
number $N_p$ (the population number $N_p$ is connected with first exponential
function in \re{a12} and was  also introduced in \ci{1}). The very large
values of $\tau$ give a dramatical increasing
of $N_p$ that leads to the instability of $E_{SIM}$ and whole stochastic
process. So, we have to choose optimal time step $\tau_0$.

CDMC \ci{AMN1} has an artifact that the averaged real quantities
become complex, but real parts of them are essentially bigger than imaginary
parts because the imaginary parts are controlled by
the phase of the Green function $\gamma_G$ depending as  $\tau_{0}^{0.5}$.
Then the imaginary part of $E_{SIM}$ for almost all number of fermions $N$
is less than 0.15 percent of its real part. For $\overline r$ and
$\overline {r^2}$ they are less than 0.07 percent of their real parts.

Table 1 present the exact analytical result for the ground
state energy of the fermionic systems in the $2D$ parabolic well $E_{EXACT}$,
 $E_{SIM}$ (in $\hbar\omega_o$ units) and the simulated values
of $\overline r$ and $\overline {r^2}$
( in $r_o$ and $r_o^2$ units respectively).
The values of $E_{EXACT}$ was calculated
by occupying the available lowest states.
The energy  of $K$-th
quantum number state of the $2D$
oscillator  is $E_K=\hbar\omega_o(K+1)$
and degeneracy number of this state is equal $(K+1)$ \ci{8}.

$E_{SIM}$ is depended on value of  numerical parameter $\alpha$
in the trial wave function $\Psi_T^*(\vec r_i)$ \re{14}.
We take $\alpha$ providing minimal value of the energy $E_{SIM}$.
The numerical data of $\alpha$ that give a minimal energy
$E_{SIM}$ are outlined also in the Table 1.
We have compared the CDMC dependence of the parameter $\alpha$ versus
number of particles
$N$ with the variational calculation of $\alpha$ \ci{18}.
If at the changing of $N$ from two to ten CDMC $\alpha$ changes from -0.593
to -0.084, so variational  $\alpha$ changes from -0.293 to -0.683
(see \ci{18}). For the big number fermions (as one can see from table)
CDMC $\alpha$  is almost the same for all systems.
Thus a mean field approximation of  Fetter, Hanna and Laghlin \ci{7} with
numerical parameter in a wave function can be considered as a good
start  approximation  for the simulation of a big number anyons.
Table 1 also shows the numerical values of the population number $N_p$
and optimal time steps $\tau_0$.
Good agreement of the simulation results for the $E_{SIM}$ with the
exact ground state energies $E_{EXACT}$ for the numbers of the
particles bigger than four allows
us to hope that CDMC \ci{AMN1} gives a possibility for the exact simulation
of the big number $2D$ fermions.

\begin{table}
\caption{The energies of the fermion systems in $2D$ parabolic well.\\
Here
$N$ - numbers of particles, $E_{EXACT}$ - analytically calculated  ground
state energies
(in $\hbar\omega_o$ units),
$E_{SIM}$ - results of the simulation  ground state energies
 (in $\hbar\omega_o$ units),
$\overline r$ - the numerical values for the  simulated mean
radius (in $r_o$ units), $\overline {r^2}$ - the numerical values for the
simulated mean square radius
(in $r_o^2$ units), $\alpha$ - the numerical parameters in the trial wave
function \re{14} that gives a minimum $E_{SIM}$, $N_p$ - the population
numbers and $\tau_0$ - the optimal time steps. All
simulated quantities and their deviations from mean values were averaged
over 30 last time blocks $\Delta t$ when $E_{SIM}$ and $N_p$ have had
relatively stable values. }

\begin {center}
\begin{tabular}{|c|c|c|c|c|c|c|c|}\hline

    N      & $E_{EXACT}$ & $E_{SIM}$ & $\overline{r}$ & $\overline{r^2}$ &
$\alpha$   & $\overline{N_p}$&$\tau_0$ \\ \hline
    2      &        3    &  1.842                &       0.935    &   0.8852
& -0.593 & $980\pm5$&0.08 \\
           &             & $\pm1.297\cdot10^{-3}$ & $\pm6.994\cdot10^{-4}$
& $\pm1.319\cdot10^{-3}$  &             & &\\ \hline
    3      &        5    &     4.399             &       1.092    & 1.573
& -0.423     & $925\pm9$&0.06 \\
           &             & $\pm4.471\cdot10^{-3}$ & $\pm1.564\cdot10^{-3}$
& $\pm3.814\cdot10^{-3}$& & & \\ \hline
    4      &        8    & 7.399                 & 1.236
& 1.919    &- 0.3        & $908\pm13$&0.04 \\
   {}      &        {}   &{$\pm4.092\cdot10^{-2}$}&{$\pm1.844\cdot10^{-2}$}
& {$\pm4.562\cdot10^{-2}$} &  &  & \\ \hline
{5}        &   {11}      &  {11.195}             &{1.452}
& {2.614}  & {-0.453}     & {$949\pm10$}&0.04 \\
   {}      &     {}      &{$\pm1.798\cdot10^{-2}$}&{$\pm3.738\cdot10^{-3}$}
&{$\pm1.324\cdot10^{-2}$} &   {}                  &{}& \\ \hline
  {6}      &    {14}     &   {14.492}            &    {1.619}
& {3.275}  &  {-0.092}    &  {$992\pm24$}&0.02 \\
   {}      &     {}      &{$\pm3.136\cdot10^{-2}$}&{$\pm4.261\cdot10^{-3}$}
&{$\pm1.564\cdot10^{-2}$} &    {}                 &{} & \\ \hline
  {7}      &   {18}      &     {18.446}          &   {1.752}
& {3.855}  &  {-0.022}    &   {$1042\pm53$}&0.02 \\
    {}     &    {}       &{$\pm9.126\cdot10^{-2}$}&{$\pm8.277\cdot10^{-3}$}
&{$\pm3.665\cdot10^{-2}$} &      {}               &{} & \\ \hline
    {8}    &    {22}     &        {22.383}       &   {2.020}
& {5.112}  & {-0.046}     &  {$1110\pm57$}&0.015 \\
   {}      &    {}       &{$\pm8.569\cdot10^{-2}$}&{$\pm5.821\cdot10^{-3}$}
&{$\pm2.790\cdot10^{-2}$} &        {}             & {} & \\ \hline
    {9}    &     {26}    &        {26.556}       &{2.294}
& {6.672}  &   {-0.067}    & {$1090\pm33$}& 0.01 \\
    {}     &      {}     &{$\pm9.595\cdot10^{-2}$}&{$\pm1.061\cdot10^{-2}$}
&{$\pm6.435\cdot10^{-2}$} &           {}          &{} &\\ \hline
    {10}   &     {30}    &        {30.905}       & {2.601}
&  {8.599} &    {-0.084}   &    {$1032\pm23$}& 0.005\\
     {}    &    {}       &{$\pm2.010\cdot10^{-1}$}&{$\pm2.839\cdot10^{-2}$}
&{$\pm2.095\cdot10^{-1}$} &          {}           & {}& \\ \hline
\end{tabular}
\end{center}
\end{table}


\begin{thebibliography}{17}

\bibitem{9}
D.M. Ceperley and M.H. Kalos in {\it Monte Carlo Methods in
Statistical Physics}, ed. by K.Binder (Springer, Verlag,
Berlin, Heidelberg, New York, 1979).

\bibitem{10}
B.O. Kerbikov, M.I. Polikarpov, L.V. Shevchenko and A.B. Zamolodchikov,
Preprint ITEP 86-160, Moscow, 1986.

\bibitem{11}
G. Ortiz, D.M.Ceperley and R.M.Martin,{\it Phys. Rev. Lett.,} {\bf 71},
2777(1993).

\bibitem{12}
L.Zhang, G.Canright and T.Barnes,{\it Phys. Rev.} {\bf B49}, 12355(1994).

\bibitem{AMN1}
B. Abdullaev, M. Musakhanov and A. Nakamura,
Complex Diffusion Monte-Carlo Method for the
 systems with complex wave function:
test by the simulation of 2$D$ electron in uniform magnetic field,
cond-mat/0101330.

\bibitem{13}
M.I.Polikarpov and L.V.Shevchenko,
Modified Monte-Carlo method for Green function...,
Preprint ITEP 43, Moscow, 1987.

\bibitem{14}
R.B.Laughlin,{\it Phys. Rev. Lett.,} {\bf 50}, 1395(1983).

\bibitem{1}
P.J. Renolds, D.M. Ceperley, B.J. Alder and W.A. Lester Jr.,
{\it J.Chem.Phys.,} {\bf 77}, 5593 (1982).


\bibitem{15}
K.E. Schmidt and M.H. Kalos in {\it Monte Carlo Methods in
Statistical Physics II}, ed. by K.Binder (Springer,
Berlin, 1984); K.E. Schmidt and D.M. Ceperley in {\it The
Monte Carlo Method in Condensed Matter Physics}, ed. by K.Binder
(Springer, Berlin,  1991).


\bibitem{16}
S. Zhang, J. Carlson and J.E. Gubernatis, {\it Phys. Rev. Lett.},
{\bf 74}, 3652 (1995); {\it Phys. Rev.} {\bf B55}, 7464 (1997).

\bibitem{17}
J. Carlson, J.E. Gubernatis, G. Ortiz and S. Zhang, cond-mat/9901029.


\bibitem{3}
F. Wilczek, {\it Phys.Rev.Lett.,} {\bf 48}, 1144 (1982); {\bf 49}, 957 (1982).

\bibitem{4}
Y.S. Wu,  {\it Phys.Rev.Lett.,} {\bf 53}, 111 (1984).

\bibitem{5}
R.B. Laughlin, {\it Phys.Rev.Lett.,} {\bf 60}, 2677 (1988).

\bibitem{7}
A.L. Fetter, C.B. Hanna and R.B. Laughlin,
{\it Phys.Rev.,} {\bf B39}, 9679 (1989).

\bibitem{8}
V.M. Galicki, B.M. Karnakov and V.I.Kogan,
{\it The problems on the quantum mechanics} (in russian),
Moscow, "Nauka", 1981, 648.

\bibitem{18}
B. Abdullaev, M. Musakhanov and A. Nakamura,
Approximate formula for the ground state energy of anyons
in $2D$ parabolic well, cond-mat/0012423.

\end{thebibliography}
\end{document}